\documentclass[reqno,11pt]{article}
\usepackage{authblk}
\usepackage{vmargin}
\setmarginsrb{2.4cm}{2.4cm}{2.4cm}{2.4cm}{0.2cm}{0.2cm}{0.2cm}{0.5cm}
\usepackage{graphicx}
\usepackage{epstopdf}
\usepackage{subfigure}
\usepackage{caption}
\usepackage{amssymb}
\usepackage{amsthm}
\usepackage{amssymb}
\usepackage{mathtools}
\usepackage{amsmath}
\usepackage{cite}
\usepackage{appendix}
\usepackage{hyperref}
\hypersetup{
    colorlinks=true,
    linkcolor=blue,
    filecolor=magenta,
    urlcolor=cyan,
    citecolor=blue,
}
\makeatletter
\newcommand*{\rom}[1]{\expandafter\@slowromancap\romannumeral #1@}
\makeatother
\newtheorem{thm}{Theorem}

\newtheorem{lemma}{Lemma}

\title{Discrete rogue waves and blow-up from solitons of a nonisospectral
semi-discrete nonlinear Schr\"{o}dinger equation}

\author{Abdselam Silem, ~~Hua Wu, ~~Da-jun Zhang\footnote{Corresponding author. Email: djzhang@staff.shu.edu.cn}}

\affil{\small  Department of Mathematics, Shanghai University, Shanghai 200444, P.R. China}

\date{\today}

\begin{document}

\maketitle

\begin{abstract}
We investigate the nonisospectral effects of a semi-discrete nonlinear Schr\"{o}dinger equation,
which is a direct integrable discretisation of its continuous counterpart.
Bilinear form and double casoratian solution of the equation are presented.
Dynamics of solutions are analyzed. Both solitons and multiple pole solutions admit space-time localized rogue wave behavior. And more interestingly, the solutions allow blow-up at finite time $t$.
\end{abstract}


\section{Introduction}\label{sec-1}

Over the past years rogue waves have been a hot topic in mathematical studies,
meanwhile had large physical applications (cf.\cite{Rogue-Nature-2007,Akhmediev2010,Onorato2013}).
Rogue waves are characterized by space-time localness \cite{AkhAT-PLA-2009}
and usually are described by rational solutions of the nonlinear Schr\"odinger equation (NLS)
\cite{Peregrine,Eleoneski1986,Akhmediev2009,Guo2012}
and by interactions of solitons (e.g. \cite{SluP-PRL-2016}).

Apart from rational solutions and multi-solition-interaction, it is possible to
obtain localized waves from a single soliton of an integrable equation with nonisospectral effects.
In fact, usually the amplitude of a soliton is governed by the spectral parameter $\lambda$
of its spectral problem.
$\lambda$  is constant in isospectral case, which leads to a usual soliton,
and time-dependent in nonsospectral case, which leads to a soliton with a time-dependent amplitude.
In particular, when $\lambda_t=\lambda^2$, which indicates $\lambda=-1/(t+c)$ where $c$ is a constant,
it yields an amplitude that decays when $|t|\to +\infty$.
Such a mechanism to generate   space-time localized waves has been
demonstrated by  a nonisospectral Mavakov system \cite{Kou-MT-2011},
nonisospectral Kadomtsev-Petviashvili(I) equation \cite{DJZ-Lump}
and nonispsoectral NLS \cite{Silem-Zhang-2019}.
Note that nonisospectral integrable systems are not only used to describe solitary waves
in nonuniform  media (cf.\cite{Chen-Liu-1976,HS-JPSJ-1976,ZhaBH-JPA-2006}),
but also related to some physics models
(e.g. with external potentials)
by transformations (cf.\cite{SHB-PRL-2007,HeL-SAPM-2010,ZhaZL-AP-2014,Liu-ROMP-2020}).

In this paper, we will investigate nonisospectral effects in discrete case and consider a semi-discrete nonisospectral NLS (sdnNLS)
\begin{equation}
\begin{array}{l}
iQ_{n,t}=\frac{1}{2}(1+Q_nQ^*_n)\left[(2n+2)Q_{n+1}+(2n-2)Q_{n-1}\right]-2nQ_n\\
 \hspace{1.3cm}+Q_n(E-1)^{-1}(Q_{n+1}Q_n^*+Q_nQ^*_{n+1}),
\end{array}
\label{New-NDnls}
\end{equation}
where $*$ stands for the complex conjugate and $E$ is the shift operator in $n$ direction.
This equation is a direct integrable discretization of the  nonisospectral NLS \eqref{nNLS-III}
that allows localized solitons.
It is interesting  that this equation not only admits discrete rogue waves as its continuous counterpart does,
but also allows a localized soliton that blows up at finite time $t$. This is different from the continuous
 case (cf.\cite{Silem-Zhang-2019}) and not reported before.
Note that as an integrable discretization, \eqref{New-NDnls} can be potentially used as
a numerical scheme of its continuous (nonautonomous) counterpart.

The paper is organized as follows. In Sec.\ref{sec-2} we provide integrable background of the sdnNLS
\eqref{New-NDnls}. In Sec.\ref{sec-3} we derive double Casoratian solutions of the equation,
and in Sec.\ref{sec-4} we illustrate its localized and blow-up property of solitons.
Finally, conclusions are presented in Sec.\ref{sec-5}.

\section{Integrability of the sdnNLS}\label{sec-2}

The sdnNLS \eqref{New-NDnls} is a result of the reduction $R_n=-Q^*_n$ of the following coupled system
\begin{equation}
\begin{array}{l}
iQ_{n,t}=\frac{1}{2}(1- Q_nR_n)\left[(2n+2)Q_{n+1}+(2n-2)Q_{n-1}\right]-2nQ_n\\
 \hspace{1.3cm}- Q_n(E-1)^{-1}\left(Q_{n+1}R_n+Q_nR_{n+1}\right),\\
iR_{n,t}=-\frac{1}{2}(1- Q_nR_n)\left[(2n+2)R_{n+1}+(2n-2)R_{n-1}\right]+2nR_n\\
 \hspace{1.3cm}+R_n(E-1)^{-1}\left(Q_{n+1}R_n+Q_nR_{n+1}\right).
\end{array}
\label{New-NDnls-s}
\end{equation}
Here, $Q_n, R_n$ are functions of $(n,t)$ defined on $\mathbb{Z}\times \mathbb{R}$,
$E$ is a shift operator defined as $E^j Q_n=Q_{n+j}$.
This equation is related to the Ablowitz-Ladik (AL) spectral problem \cite{AL-1975,AL-1976}
and have the following Lax Pair \cite{Zhang-2010}
\begin{equation}
\label{lax.1}
\Phi_{n+1}=M_n\Phi_n,\ \ \Phi_{n,t}= {N}\Phi_n,~~
M_n=\begin{pmatrix}
\lambda & Q_n\\
R_n & \frac{1}{\lambda}
\end{pmatrix},
\ \  {N}=\begin{pmatrix}
 {A}_n &  {B}_n\\
 {C}_n &  {D}_n
\end{pmatrix},
~~\ \ \Phi_n=\begin{pmatrix}
\phi_{1,n}\\
\phi_{2,n}
\end{pmatrix},
\end{equation}
where $\lambda_t=\lambda^3-2 \lambda+\lambda^{-1}$ (or alternatively, $z_t=2z^2$ if $z=\lambda^2-1$),
\[\begin{array}{l}
A_n=-i[(n-\frac{1}{2})(\lambda^2-2+\lambda^{-2})-(2n-2)Q_nR_{n-1}-2(E-1)^{-1}Q_{n+1}R_n], \\
B_n=-i[-(2n-2)Q_{n-1}\lambda^{-1}+2nQ_n\lambda],~~
C_n=i[2nR_n\lambda^{-1}-(2n-2)R_{n-1}\lambda],\\
D_n=i[(n-\frac{1}{2})(\lambda^2-2+\lambda^{-2})-(2n-2)Q_{n-1}R_n-2(E-1)^{-1}Q_nR_{n+1}].
\end{array}
\]

For the sdnNLS \eqref{New-NDnls}, defining $x=nh$ and
replacing $Q_n\to h q(x,t)$, $t\to t/h^2$,  then when $n\to\infty,~ h\to 0$,
the leading term of $h$ in \eqref{New-NDnls} yields
the continuous nonisospectral NLS (nNLS) \cite{Zhang-2010}
\begin{equation}\label{nNLS-III}
iq_{t}+x(q_{xx}+ 2|q|^2q )+4q_{x}+q \partial^{-1}|q|^2=0,
\end{equation}
where $\partial^{-1}\partial_x=\partial_x\partial^{-1}=1$,
which has demonstrated that rogue waves can arise from solitons with nonisospectral effects \cite{Silem-Zhang-2019}.

\section{Bilinear Form and Casoartian solutions}\label{sec-3}

Our strategy is  first solving  the unreduced  coupled system \eqref{New-NDnls-s}
and then imposing reduction on solutions to get solutions to the sdnNLS \eqref{New-NDnls}.

By the dependent variables transformation
\begin{equation}
\label{H-T}
Q_n=\frac{g_n}{f_n},~ R_n=\frac{h_n}{f_n},
\end{equation}
we obtain the following  bilinear form of \eqref{New-NDnls-s}:
\begin{subequations}\label{Gblf}
\begin{align}
&iD_tg_n.f_n =\frac{1}{2}
[(2n+2)g_{n+1}f_{n-1}+(2n-2)g_{n-1}f_{n+1}-4ng_nf_n] +g_nv_n,\label{blf-1.1} \\
&iD_th_n.f_n =-\frac{1}{2}
[(2n+2)h_{n+1}f_{n-1}+(2n-2)h_{n-1}f_{n+1}-4nh_nf_n] -h_nv_n,\label{blf-1.2} \\
& f_n^2-f_{n-1}f_{n+1} = g_nh_n, \label{blf-2}\\
& v_{n+1}f_n-v_nf_{n+1} = -g_{n+1}h_n-g_nh_{n+1} \label{blf-3},
\end{align}
\end{subequations}
where $D$ is the well known Hirota bilinear operator defined in \cite{Hirota-1974} as\\
\begin{equation*}
 D_t^j\ f\cdot g\equiv (\partial_t-\partial_{t^{\prime}})^j f(t) g(t^{\prime})|_{t^{\prime}=t}.
 \end{equation*}
 and $v_n$ is an auxiliary function to be defined later.

Consider matrix equations
\begin{subequations}\label{Cas-cnd}
\begin{align}
\label{Phi-cnd}
&E\Phi_n=A\Phi_n,~~i\Phi_{n,t}=\frac{n}{2}\left(E^2-2+E^{-2}\right)\Phi_n-N(E^2-2)\Phi_n,
\\
\label{Psi-cnd}
&E^{-1}\Psi_n=A\Psi_n,~~i\Psi_{n,t}=-\frac{n}{2}\left(E^2-2+E^{-2}\right)\Psi_n+M(E^2-2)\Psi_n,
\end{align}
\end{subequations}
where $A\in \mathbb{C}_{(N+M+2)\times (N+M+2)}$ is an invertible matrix, and
\begin{align*}
\Phi_n=(\phi_{1,n},\phi_{2,n},\ldots,\phi_{N+M+2,n})^T,~ \Psi_n=(\psi_{1,n},\psi_{2,n},\ldots,\psi_{N+M+2,n})^T
\end{align*}
are $(N+M+2)$-th order vectors, for which the each component $\phi_{j,n}$,
$\psi_{j,n}$ is a function of  $n$ and $t$.
Introduce  double Casoratian (cf.\cite{Deng-Zhang-2018})
\begin{equation}
\begin{array}{rl}
\mathrm{Cas}^{(N+1,M+1)}(\Phi_n,\Psi_n)
=&|\Phi_n,E^2\Phi_n,\ldots,E^{2N}\Phi_n;\Psi_n,E^2\Psi_n,\ldots,E^{2M}\Psi_n|
=|\widehat{\Phi_n^{(N)}};\widehat{\Psi_n^{(M)}}|\\
=&|0,1,\ldots,N;0,1,\ldots,M|=|\widehat{N};\widehat{M}|,
\end{array}
\end{equation}
where the notation ``hat" is employed for the compact expression (cf.\cite{Nimmo-NLS}).
Then,  for solutions to Eq. \eqref{New-NDnls-s} as well to the bilinear form \eqref{Gblf}, we have the following.

\begin{thm}\label{thm-1}
The bilinear form \eqref{Gblf} has double Casoration solutions
\begin{subequations}\label{cnd-sol}
\begin{align}
&f_n=|\widehat{N};\widehat{M}|, ~g_n=|\widehat{N+1};\widehat{M-1}|, ~h_n=-|\widehat{N-1};\widehat{M+1}|,\\
&v_n=|\widehat{N};\widehat{M-1},M+1|+|\widehat{N-1},N+1;\widehat{M}|-|\widehat{N};\widehat{M}|,
\end{align}
\end{subequations}
where the entries $\Phi_n$ and $\Psi_n$ satisfy the conditions \eqref{Cas-cnd}.

\end{thm}
Proof   will be sketched later in the appendix.

 Such  $\Phi_n$ and $\Psi_n$ can be constructed by assuming
\begin{equation}
\label{new-phi-psi-0}
\Phi_n=W^N A^{n}C^{+},~~\Psi_n=W^{-M} A^{-n}C^{-},
\end{equation}
where $C^{\pm}$ are $\mathbb{C}_{N+M+2}$ column vectors,
$A, W \in \mathbb{C}_{(N+M+2)\times (N+M+2)}[t]$ are matrix functions of $t$,
$A, W, A_t$ and $W_t$ commute with each other and satisfy
\begin{equation}\label{AW}
A_t=-\frac{i}{2}(A^3-2A+A^{-1}),~~
W_t=i(A^2-2)W.
\end{equation}

To obtain solutions of \eqref{New-NDnls} by reduction, we take $N=M$ in \eqref{new-phi-psi-0}
and assume
\begin{equation}\label{tr-phi-psi}
\Psi_n=T\Phi_n^*,
\end{equation}
where $T\in \mathbb{C}_{(2N+2)\times(2N+2)}$ some transform matrix.
It then follows from \eqref{new-phi-psi-0} and \eqref{tr-phi-psi} that
$A, W, T$ and $C^{\pm}$ obey the following
\begin{equation}\label{A-W-cond}
A^{-1}=TA^*T^{-1},~~ W^{-1}=TW^*T^{-1}, ~~C^-=TC^{+*}.
\end{equation}
Consider one more assumption
\begin{equation}\label{T-cd}
TT^*=-I_{2(N+2)},
\end{equation}
 where $I_{2(N+2)}$ is the $2(N+2)$-th order
identity matrix. With this assumption  we then have
\[ f_n^*=|\widehat{\Phi_n^{(N)}};T\widehat{\Phi_n^{*(N)}}|^*
=|T|^{-1}|\widehat{T\Phi_n^{*(N)}};TT^*\widehat{\Phi_n^{(N)}}|=|T|^{-1}f_n\]
and similarly, $g_n^*=-|T|^{-1}h_n$.
Thus from \eqref{H-T} we immediately have $R_n=-Q^{*}$.
If taking $A=e^B$ and $W=e^{\Omega}$ where $B,~\Omega\in \mathbb{C}_{(2N+2)\times (2N+2)}[t]$,
$\Phi_n$ in \eqref{new-phi-psi-0} can be rewritten as
\begin{equation}\label{PHI}
\Phi_n=e^{nB+N\Omega}C^+,
\end{equation}
and to keep \eqref{AW} and \eqref{A-W-cond} valid it is sufficient to consider
\begin{equation}\label{BOme}
B_t=-\frac{i}{2}(e^{2B}-2+e^{-2B}),~~
\Omega_t=i(e^{2B}-2),
\end{equation}
and
\begin{equation}\label{new-con}
BT+TB^*=0,~\Omega T+T\Omega^*=0.
\end{equation}
Note that \eqref{new-con} is one of the master relation appearing in double Wronskian reductions
(cf.\cite{CheZ-AML-2018,CheDLZ-SAPM-2018}).
The above results of reduction are concluded in the following Theorem.

\begin{thm}\label{thm-2}
The sdnNLS  \eqref{New-NDnls} has solutions $Q_n=g_n/f_n$ where
\begin{equation}
\label{new-cas}
f_n=|\widehat{\Phi_n^{(N)}},T\widehat{\Phi_n^{*(N)}}|,~~
g_n=|\widehat{\Phi_n^{(N+1)}},T\widehat{\Phi_n^{*(N-1)}}|,
\end{equation}
where $B,~\Omega$ and $T$ obey the evolution \eqref{BOme} and the constraints \eqref{T-cd} and \eqref{new-con}.
\end{thm}

Next, we present solution $B, \Omega, T$ and then $\Phi_n$.
Assume $2\times 2$ block matrices
\begin{equation}
T=\begin{pmatrix}
0 & I_{\hbox{\tiny \it{N+1}}}\\
-I_{\hbox{\tiny \it{N+1}}}& 0
\end{pmatrix},~~
B=\begin{pmatrix}
K & 0\\
0 & -K^*
\end{pmatrix},
~~
\Omega=\begin{pmatrix}
W_1 & 0\\
0 & -W_1^*
\end{pmatrix},
\end{equation}
where $K, W_1\in \mathbb{C}_{(N+1)\times (N+1)}[t]$.
The first case is both $K$ and $W_1$ are diagonal, i.e.
\begin{equation}
\label{K-form}
K=\mathrm{Diag}\left[k_0(t),k_1(t),\cdots, k^{}_{N}(t)\right],
~~
W_1=\mathrm{Diag}\left[w_0(t),w_1(t),\cdots, w^{}_{N}\right],
\end{equation}
where
\begin{align}
k_j(t)=\ln(1-\frac{i}{t+\beta_j}),~~
w_j(t)=-it+\ln(t+\beta_j),
\end{align}
$\{\beta_j\} $ are distinct complex numbers.
In this case,
\begin{align}
\Phi_n=(\phi^{}_{0,n}, \phi^{}_{1,n}, \cdots, \phi^{}_{N,n};
\psi^{*}_{0,n}, \psi^{*}_{1,n}, \cdots, \psi^{*}_{N,n})^T,
\label{entry-diag-a}
\end{align}
where
\begin{subequations}\label{Phi-soliton}
\begin{equation}
\phi_{j,n}=  e^{\rho_j}, \quad
\psi_j =  e^{-\rho_j},
\end{equation}
with
\begin{equation}
\label{rho}
\rho_j=\frac{n}{2}k_j(t)+Nw_j(t)+\rho_j^{(0)},~~ \rho_j^{(0)}\in \mathbb{C}.
\end{equation}
\end{subequations}

The second case is when $K$ is a Jordan block matrix.
For convenience we introduce lower triangular Toeplitz matrix (LTTM)
\begin{equation}
\label{toep}
\mathbf{T}=\begin{pmatrix}
a_0& 0 & 0 & \ldots & 0 & 0\\
a_1 & a_0 & 0 & \ldots & 0 & 0\\
\ldots & \ldots & \ldots & \ldots & \ldots & \ldots\\
a^{}_{N-1}& a^{}_{N-2} & a^{}_{N-3} & \ldots & a_1 & a_0
\end{pmatrix}_{N\times N}.
\end{equation}
Note that all such matrices compose an Abelian group if $a_0\neq 0$.
If $a_j=\partial^j_k p(z)/j!$, we call \eqref{toep} a LTTM generated by
function $p(z)$ and denote it by $\mathbf{T}_N[p(z)]$.
Such matrices have proved being powerful in presenting multiple pole solutions
(cf.\cite{D-Zhang-KdV,Zhang-mKdv,ZZ-SAPM-2013,TZ-AML-2020}).
Using Proposition 2.3 in \cite{D-Zhang-KdV} we have $\mathbf{T}_N[p(z)q(z)]=\mathbf{T}_N[p(z)]\mathbf{T}_N[q(z)]$,
and therefore $\mathbf{T}_N[p^m(z)]=(\mathbf{T}_N[p(z)])^m$
and $\mathbf{T}_N[e^{p(z)}]=e^{\mathbf{T}_N[p(z)]}$.
When $K=\mathbf{T}_{N+1}[k(\beta_0)]$,
one can verify that \eqref{Cas-cnd} has solution
\begin{subequations}\label{PhiPsi}
\begin{equation}
\label{gen-sol-Phi}
\Phi_n=\begin{pmatrix}
\Phi^{+}\\
\Phi^{-*}
\end{pmatrix},
\end{equation}
where
\begin{equation}
\Phi^+=(\mathbf{T}_{N+1}[e^{w(\beta_0)}])^{N}(\mathbf{T}_{N+1}[e^{\frac{1}{2}k(\beta_0)}])^n H^+,~~
\Phi^-=(\mathbf{T}_{N+1}[e^{w(\beta_0)}])^{-N}(\mathbf{T}_{N+1}[e^{\frac{1}{2}k(\beta_0)}])^{-n} H^-,
\end{equation}
\end{subequations}
and $H^{\pm}=(h_0^{\pm},h_1^\pm, \cdots, h_N^\pm)^T\in \mathbb{C}_{N+1}$.

One can also consider mixed case where
\begin{equation}
\label{mix-sol}
B=\mathrm{Diag}(\mathbf{T}_{N_0}[k(\beta_0)],\mathrm{T}_{N_1}[k(\beta_1)],\cdots,
\mathbf{T}_{N_0}[k(\beta_s)],\mathrm{Diag}[k_{s+1},k_{s+2},\ldots,k_{s+m}]),
\end{equation}
where for $m+\sum_{j=0}^s N_j =N+1$.
In this case $\Phi_n$ can be composed accordingly using the elements obtained in the first two cases.

\section{Rogue waves and blow-up of solitons }\label{sec-4}

\begin{figure}[h]
\vskip -10pt
\centering
\subfigure[ ]{
\begin{minipage}{.27\textwidth}
\centering
\includegraphics[width=5cm]{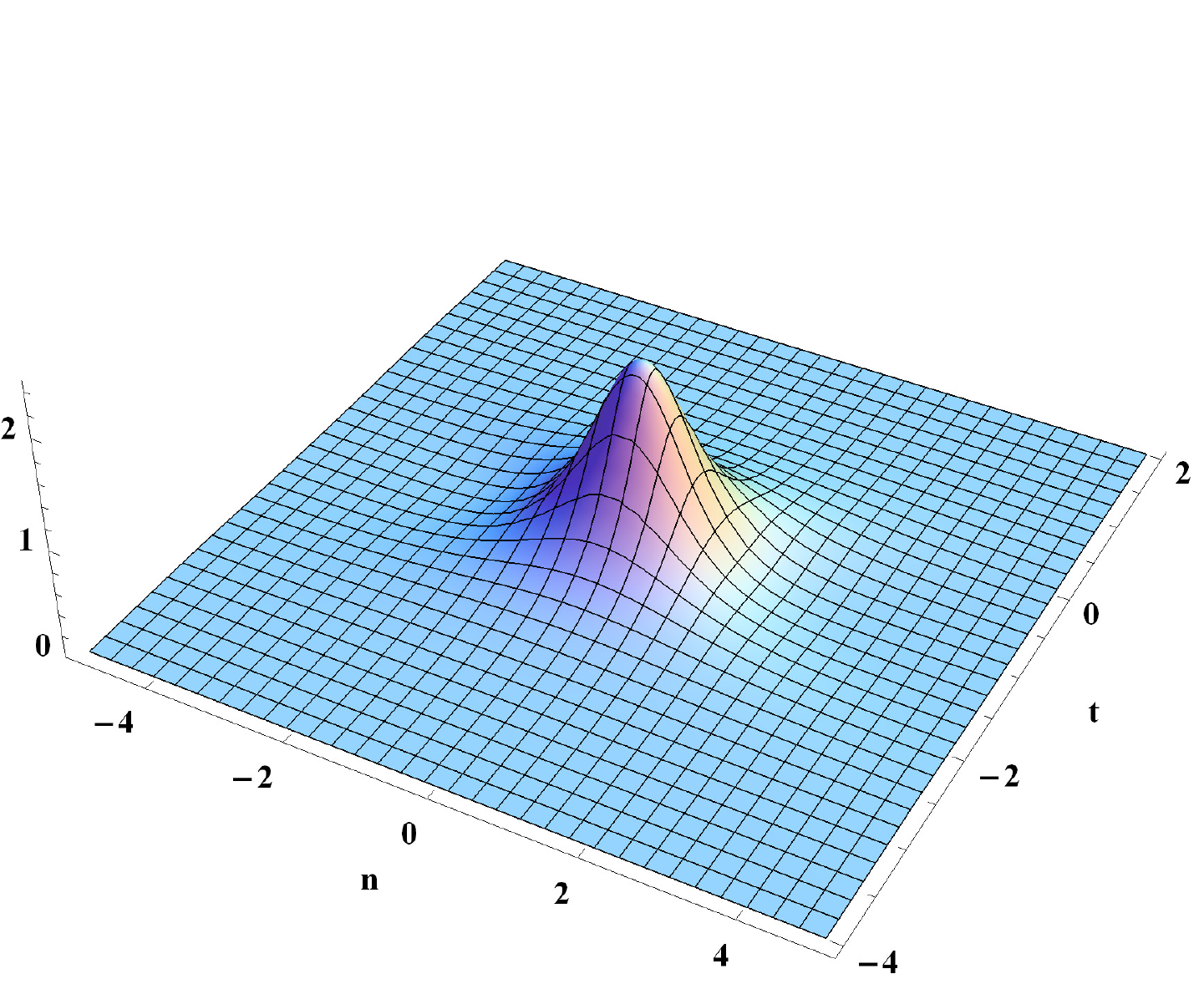}
\end{minipage}%
}
\hspace{2em}
\centering
\subfigure[ ]{
\begin{minipage}{0.3\textwidth}
\centering
\includegraphics[width=5cm]{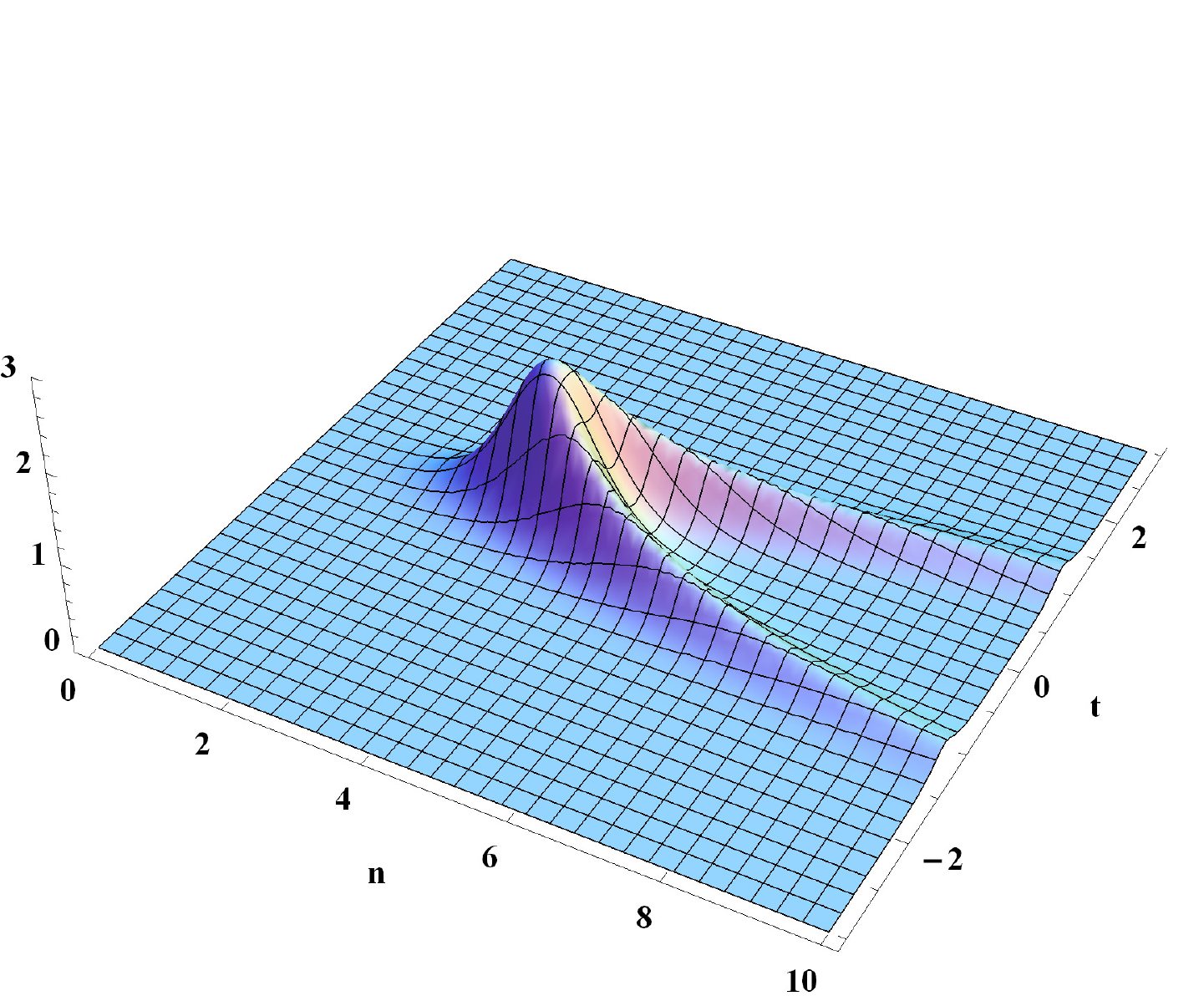}
\end{minipage}
}%
\subfigure[ ]{
\begin{minipage}{0.27\textwidth}
\centering
\includegraphics[width=5cm]{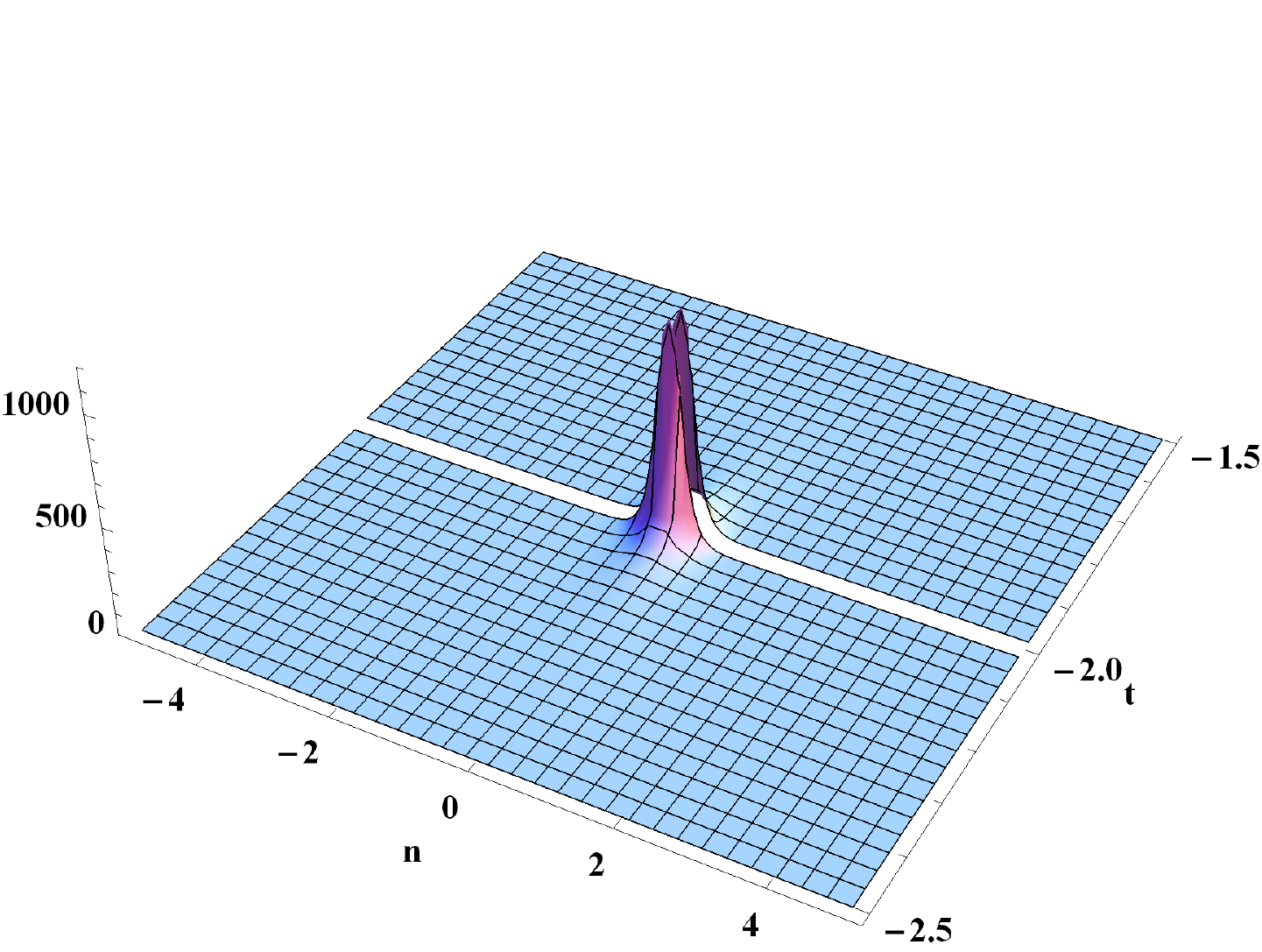}
\end{minipage}
}%
\caption{Shape and motion of 1-soliton \eqref{1-ss} of the sdnNLS \eqref{New-NDnls}:
(a) Envelope of stationary solitary wave  for $\beta_0=1+1.5 i, \rho_0^{(0)}=0$;
(b) envelope of a moving solitary wave for $\beta_0=1.5i, \rho_0^{(0)}=2$;
(c) blow-up of rogue wave for $\beta_0=2+i, \rho_0^{(0)}=0$.}
\label{fig1}
\end{figure}

From now on, in the expression of $\rho_j$ in Eq. \eqref{rho}, we take
$\beta_j=a_j+ib_j$ and $\rho_j^{(0)}=c_j+id_j$, where $a_j, b_j, c_j, d_j\in \mathbb{R}$
and $b_j\neq \frac{1}{2}$. One should notice that from Eq.\eqref{blf-2}
and through the reduction $R_n=-Q_n^*$, we deduce that
\begin{equation}
\label{Qns}
|Q_n|^2=\frac{f_{n-1}f_{n+1}-f_n^2}{f_n^2}.
\end{equation}
Thus, when $N=0$  and employing  \eqref{new-cas} and \eqref{Phi-soliton}, $1$-soliton solution is given as
\begin{equation}
\label{1-ss}
|Q_n|^2= \Lambda \, \mathrm{sech}^2\left[\frac{n}{2}H(t)+2c_0\right],
\end{equation}
where
\begin{equation}
\label{amp}
H(t)=\ln\left[\frac{(t+a_0)^2+(b_0-1)^2}{(t+a_0)^2+b_0^2}\right],~~
\Lambda=\frac{(1-2b_0)^2}{4[(t+a_0)^2+b_0^2][(t+a_0)^2+(b_0-1)^2]}.
\end{equation}
Apparently $\Lambda$ can serve as a time-varying amplitude due to the nonisospectral properties of
Eq.\eqref{New-NDnls}, and decays to zero as $t\rightarrow \pm \infty$.
By calculation one can find that the vertex of \eqref{1-ss} takes place at
\begin{equation}\label{ver}
(\tilde{t},\tilde{n})=\left(-a_0,\frac{-2c_0}{\ln |b_0-1|-\ln |b_0|}\right),
\end{equation}
where the maximum value of $|Q_n|^2$ is
$M_{1ss}=(1 - 2 b_0)^2/(4b_0^2(b_0-1)^2)$.
This indicates that the soliton is stationary when $c_0=0$ and $M_{1ss}$ can be arbitrarily large
when $|b_0|$ is small enough.
Thus $|Q_n|^2$ \eqref{1-ss} can provide a localized (rogue) wave with arbitrarily high amplitude,
which is different from the well known the rogue wave presented by rational solution of the NLS
(cf.\cite{Peregrine}).
And more interestingly, when $b_0=0$ or $1$, the rogue wave will blow up at  $t=-a_0$.
Note that in continuous case (cf.\cite{Silem-Zhang-2019}) the nonisospectral NLS \eqref{nNLS-III}
does not have blow-up behavior.
Fig.\ref{fig1} shows stationary,  moving normal rogue waves and blow-up soliton.

\begin{figure}[h!]
\vskip -10pt
\subfigure[ ]{
\begin{minipage}{.27\textwidth}
\centering
\includegraphics[width=5cm]{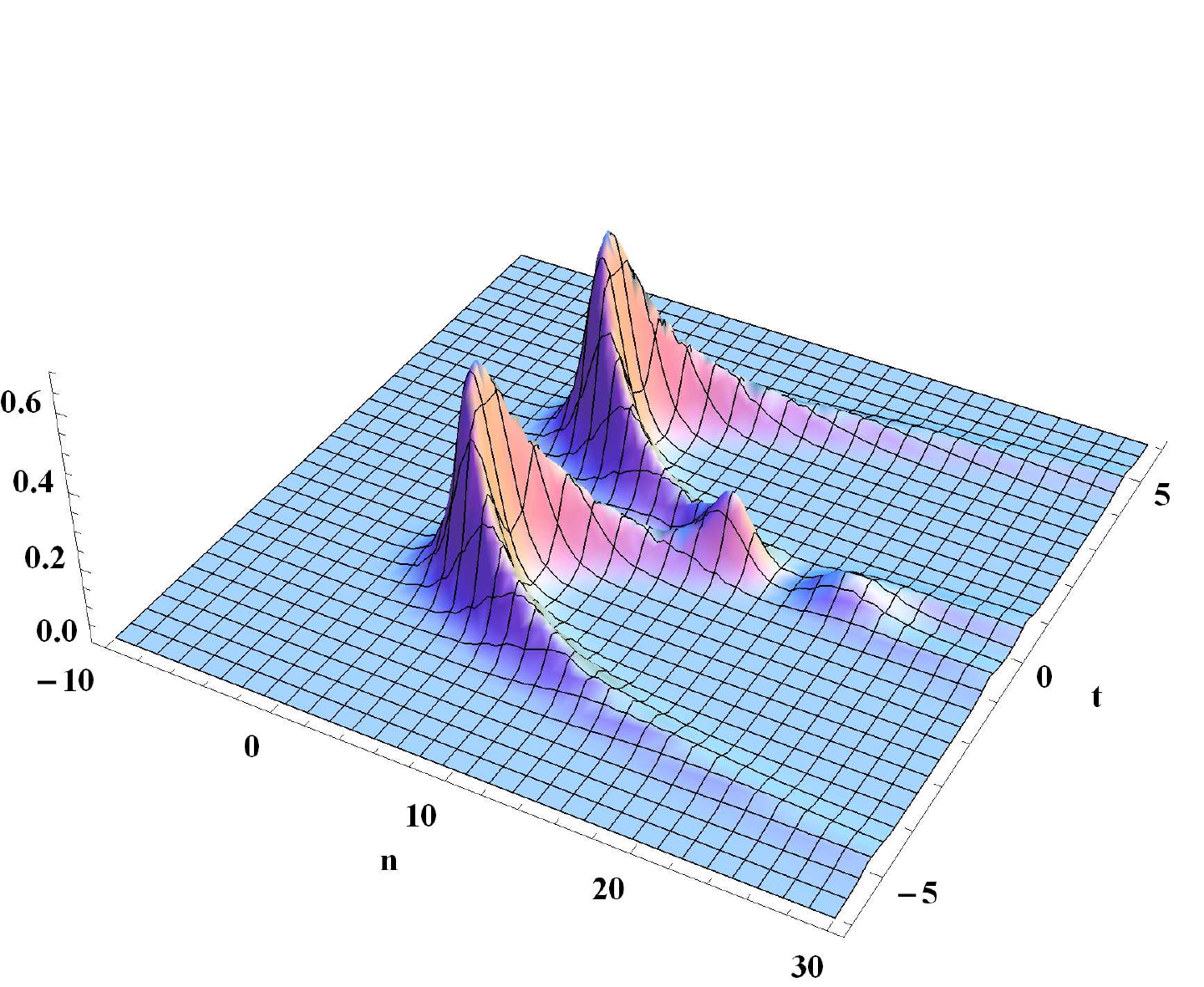}
\end{minipage}%
}
\hspace{1.5em}
\centering
\subfigure[ ]{
\begin{minipage}{0.3\textwidth}
\centering
\includegraphics[width=5cm]{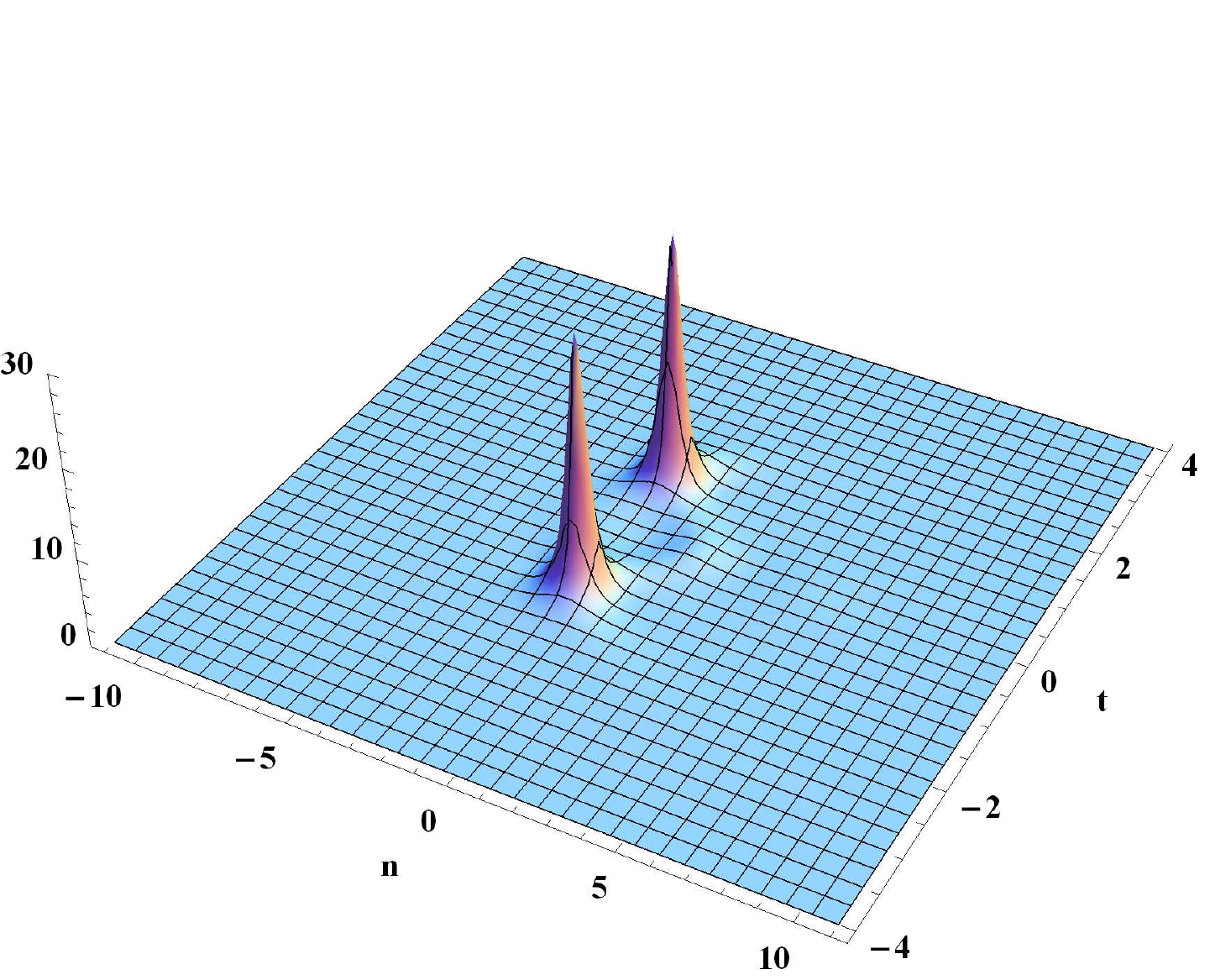}
\end{minipage}
}%
\subfigure[ ]{
\begin{minipage}{0.27\textwidth}
\centering
\includegraphics[width=5cm]{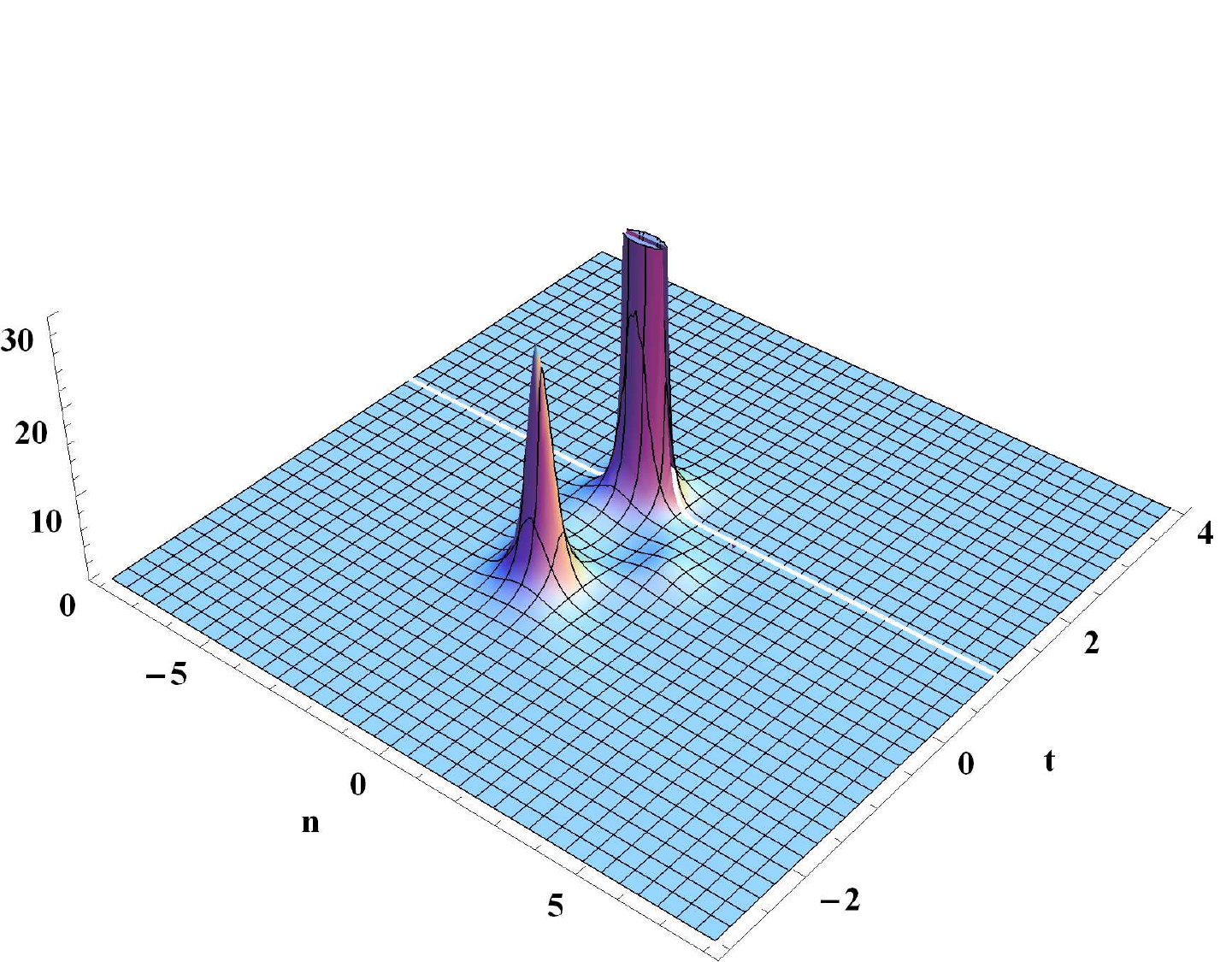}
\end{minipage}
}%
\caption{Shape and motion of 2-soliton given by \eqref{Qns} with \eqref{2-ss}:
(a) Interactions of two rogue waves  for $\beta_0 =-2+2 i$,
$\beta_1=-2+2 i, \rho_0^{(0)}=\rho_1^{(0)}=1$;
(b) two separated rogue waves for $\beta_0 =1+1.1 i$, $\beta_1=-1+1.1 i, \rho_0^{(0)}=\rho_1^{(0)}=0$;
(c) two separated rogue waves with one blow-up for $\beta_0 =0.5+1.1 i$, $\beta_1=-1+ i, \rho_0^{(0)}=\rho_1^{(0)}=0$.}
\label{fig2}
\end{figure}

2-soliton solution can be obtained by taking $N=1$ in double casoration formula, then $f_n$ in \eqref{Qns} is given by
\begin{equation}
\label{2-ss}
f_n=|\Phi_n,E^2\Phi_n;\Psi_n;E^2\Psi_n|,
\end{equation}
with
$\Phi_n=(e^{\rho_0},e^{\rho_1},e^{-\rho_0^*},e^{-\rho_1^*})^T,~ \Psi_n=T\Phi^*_n,$
where $\rho_j$ is defined in \eqref{rho}.
Fig.\ref{fig2} illustrates  two soliton solutions.

\begin{figure}[h!]
\centering
\vskip -10pt
\subfigure[ ]{
\begin{minipage}{.27\textwidth}
\centering
\includegraphics[width=5cm]{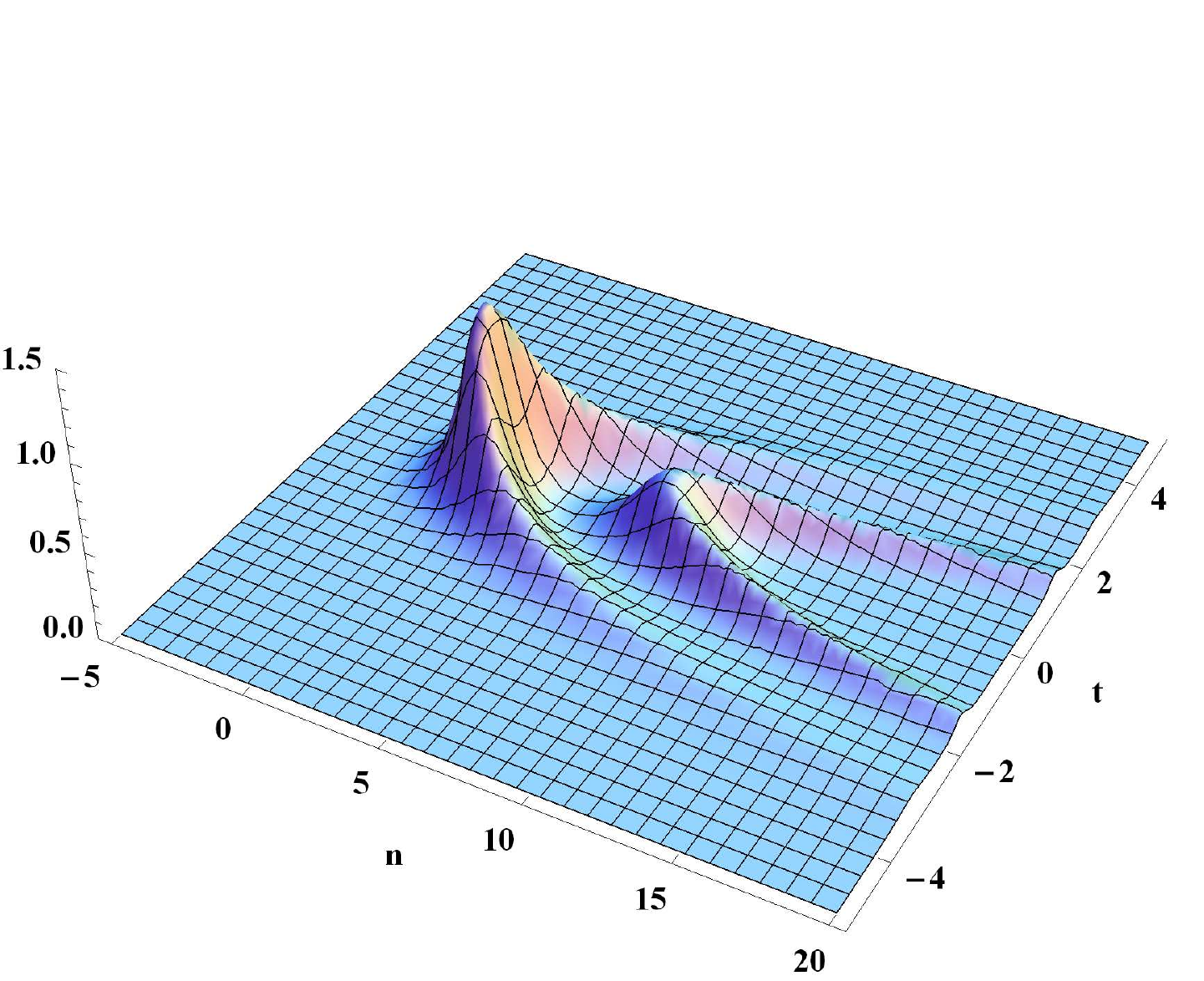}
\end{minipage}%
}
\hspace{1.5em}
\centering
\subfigure[ ]{
\begin{minipage}{0.3\textwidth}
\centering
\includegraphics[width=5cm]{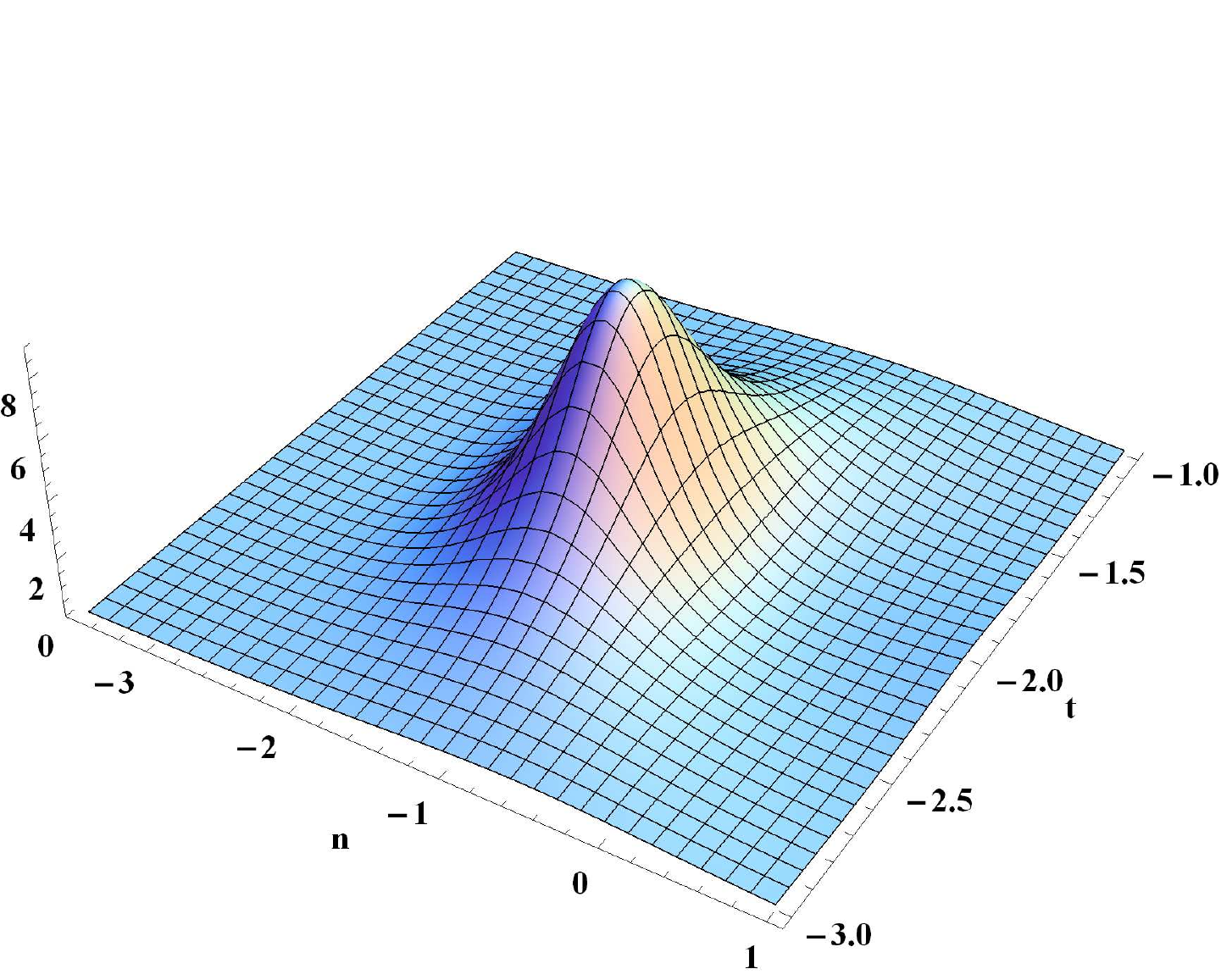}
\end{minipage}
}%
\subfigure[ ]{
\begin{minipage}{0.27\textwidth}
\centering
\includegraphics[width=5cm]{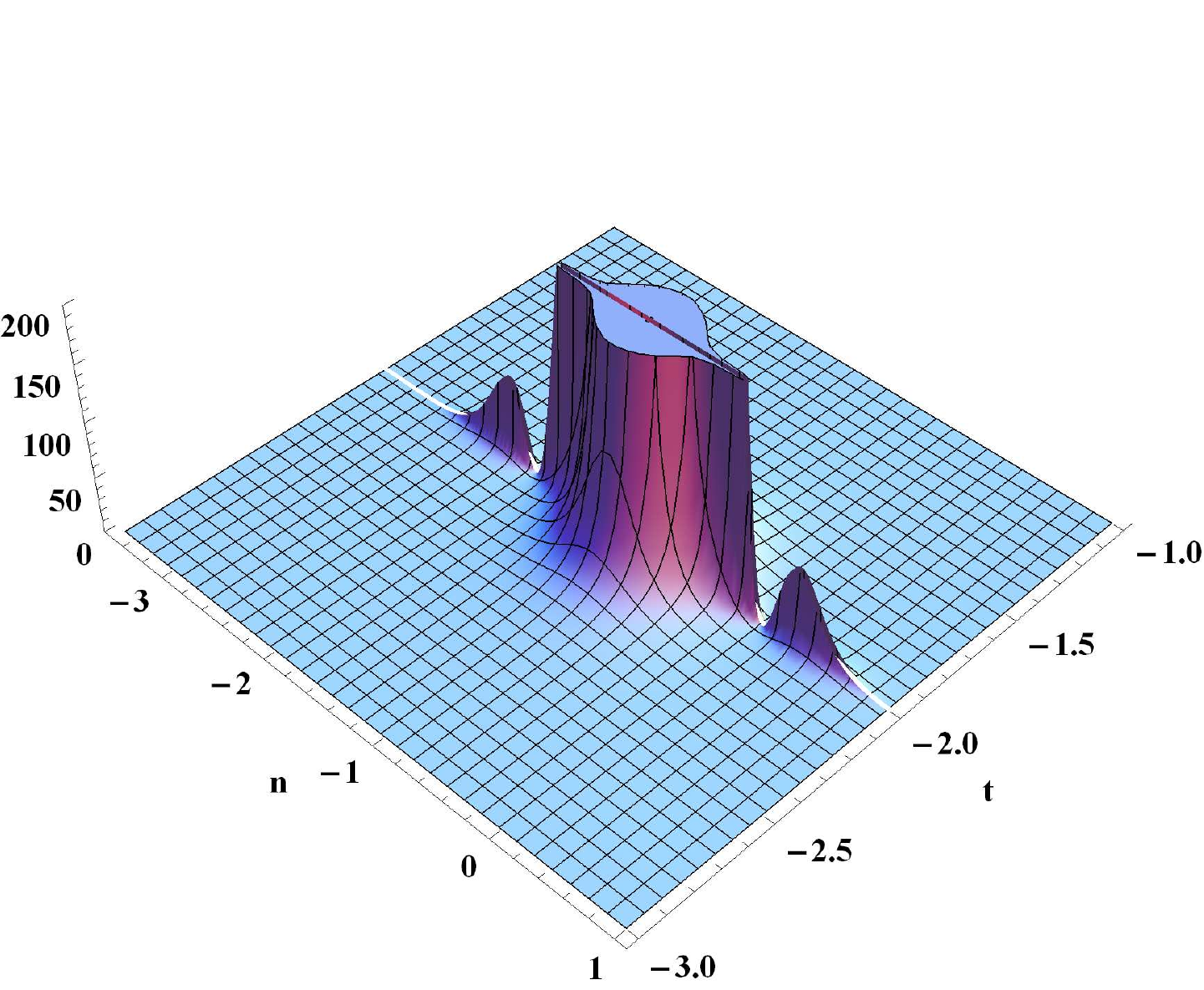}
\end{minipage}
}
\caption{Shape and motion of Jordan block solution  given by \eqref{Qns} with \eqref{2-ss} and \eqref{phi-psi-JB}:
(a) Two rogue waves for $\beta_0=2 i$ and $\rho_0^{(0)}=1$;
(b) single rogue wave for $\beta_0 =2+1.5 i$ and $\rho_0^{(0)}=0$;
(c) blow-up  of (b) for  $\beta_0=2+i$ and $\rho_0^{(0)}=0$.}
\label{fig3}
\end{figure}

Considering the following $2\times 2$ Jordan block solution by setting up in \eqref{2-ss} that
\begin{equation}
\label{phi-psi-JB}
\Phi_n=\left(e^{\rho_0},\partial_{\tiny \beta_0}e^{\rho_0},e^{-\rho_0^*},\partial_{\tiny \beta_0^*}e^{-\rho_0^*}\right)^T,
~~\Psi_n=T\Phi^*_n,
\end{equation}
where $\rho_0$ is defined as in \eqref{rho}.
This is a result of \eqref{PhiPsi} where $h^\pm_j=\delta_{j,0}\rho_0^{(0)}$.
The resulted solution is depicted in Fig.\ref{fig3}.

Note that in these cases one can also get blow-up solitons as same as the 1-soliton solution by taking $b_j\in \left\{0,1\right\}$ at  finite time $t=-a_j$ for $j=1,2$, as illustrated in the figures.

\section{Conclusion}\label{sec-5}

In this paper we derived double Casoratian solutions and investigated nonisospectral effects
of solitons of the sdNLS \eqref{New-NDnls} that is the direct integrable discretisation of nNLS \eqref{nNLS-III}.
Both solitons and multiple pole solutions admit space-time localized rogue wave behavior.
And more interestingly, the sdnNLS solutions allow blow-up at finite time $t$,
which is different from the continuous case where it has no blow-up (cf.\cite{Silem-Zhang-2019}).
Since  nonisospectral effects can be used to describe nonuniform  media (cf.\cite{Chen-Liu-1976,HS-JPSJ-1976,ZhaBH-JPA-2006}),
and  also are related to some physically significant systems (e.g. NLS with external potentials)
via transformations (cf.\cite{SHB-PRL-2007,HeL-SAPM-2010,ZhaZL-AP-2014,Liu-ROMP-2020}),
it will be interesting to investigate integrable discretisation and nonisospectral effects
of other nonisospectral integrable systems.

\vskip 10pt
\subsection*{Acknowledgments}
This project is  supported by the NSF of China (Nos.11875040 and 11631007).

\vskip 10pt
\appendixpage
\begin{appendix}
\section{Proof of Theorem \ref{thm-1}}
\begin{proof}
First of all, note that
\begin{equation}
\label{El-form}
\begin{array}{l}
E^m\Phi_{n,t}=\frac{1}{2}\left[(n+m)(E^2-2+E^{-2})\Phi_n-2N(E^2-2)\Phi_n\right],\\
E^m\Psi_{n,t}=-\frac{1}{2}\left[(n+m)(E^2-2+E^{-2})\Psi_n-2M(E^2-2)\Psi_n\right].
\end{array}
\end{equation}
In order to proceed the proof, let us recall the following lemma \cite{Freeman-1983}.
\begin{lemma}
\label{lem1}
\begin{equation}
|M,a,b|.|M,c,d|-|M,a,c|.|M,b,d|+|M,a,d|.|M,b,c|=0
\end{equation}
where M is a $N\times (N-2)$ matrix and a,b,c,d are $Nth$ order column vectors.
\end{lemma}
A direct computation yields
\[f_{n+1}=|A||\widehat{N};\widetilde{M+1}|=|A|^{-1}|\widetilde{N+1};\widehat{M}|,
~~f_{n-1}=|A|^{-1}|\widetilde{N};-1,\widehat{M-1}|,\]
\[g_{n+1}=|A||\widehat{N+1};\widetilde{M}|,~~g_{n-1}=|A|^{-1}|-1,\widehat{N};\widehat{M-1}|,\]
and  making use of \eqref{El-form} we get
\begin{equation*}
\begin{array}{rl}
if_{n,t}=&\frac{n}{2}\left(|\widehat{N-1},N+1;\widehat{M}|-|\widehat{N};-1,\widetilde{M}|
+|-1,\widetilde{N};\widehat{M}|-|\widehat{N};\widehat{M-1},M+1|+2(M-N)f_n\right)\\
&+2f_n\left(\underset{j=1}{\overset{N}{\sum}}(N-j)-\underset{j=1}{\overset{M}{\sum}}(M-j)\right),
\end{array}
\end{equation*}
\begin{equation*}
\begin{array}{rl}
ig_{n,t}=&\frac{n}{2}(|\widehat{N},N+2;\widehat{N-1}|-|\widehat{N+1};-1,\widetilde{N-1}|
+|-1,\widetilde{N+1};\widehat{N-1}|-|\widehat{N+1};\widehat{N-2},N+1|\\
&+(M-N-2)g_n)+|\widehat{N},N+2;\widehat{N-1}|+|\widehat{N+1};\widehat{N-2},N+1|\\
&+2g_n\left(\underset{j=1}{\overset{N+1}{\sum}}(N-j)-\underset{j=1}{\overset{M-1}{\sum}}(M-j)\right),
\end{array}
\end{equation*}
where we admit that $|\widetilde{N+1};\widehat{N}|=|1,2,\ldots,N+1;0,1,2\ldots,N|$.
Employing the previous identities into \eqref{blf-1.1}, one attains
\begin{equation*}
\begin{array}{rl}
&2(ig_{n,t}f_n-if_{n,t}g_n-\frac{1}{2}\left[(2n+2)g_{n+1}f_{n-1}+(2n-2)g_{n-1}f_{n+1}-4ng_nf_n\right]-g_nv_n)\\
=& n[|A|^{-2}(|\widehat{N},N+2;\widehat{M-1}||\widetilde{N+1};-1,\widehat{M-1}|
-|\widetilde{N},N+2;-1,\widehat{M-1}||\widehat{N+1};\widehat{M-1}|\\
&-|\widetilde{N+2};\widehat{M-1}||\widehat{N};-1,\widehat{M-1}|)\\
&+|A|^2(|-1,\widehat{N};\widetilde{M-1},M+1||\widehat{N};\widehat{M}|
-|\widehat{N};\widehat{M-1},M+1||-1,\widehat{N};\widehat{M}|\\
&+|\widehat{N};\widetilde{M+1}||-1,\widehat{N};\widehat{M-1}|)]\\
&+2(|A|^{-2}(|\widehat{N},N+2;\widehat{M-1}||\widetilde{N+1};-1,\widehat{M-1}|
-|\widetilde{N+2};\widehat{M-1}||\widehat{N};-1,\widehat{M-1}|)\\
&+|A|^2(|-1,\widehat{N};\widetilde{M-1},M+1||\widehat{N};\widehat{M}|
+|\widehat{N};\widetilde{M+1}||-1,\widehat{N};\widehat{M-1}|)\\
&-g_nf_n-g_{n+1}f_{n-1}+g_{n-1}f_{n+1}-g_nv_n).
\end{array}
\end{equation*}
For the $n$ coefficient, it's equal to $0$ as proved  in \cite{Deng-Zhang-2018},
and using Lemma \ref{lem1} we work out
\begin{equation}
\label{vnC}
v_n=|\widehat{N};\widehat{M-1},M+1|+|\widehat{N-1},N+1;\widehat{M}|-|\widehat{N};\widehat{M}|,
\end{equation}
and Eq.\eqref{blf-1.1} is verified.
In a similar procedure, Eq.\eqref{blf-1.2} can be verified.
For Eq.\eqref{blf-2}, see Ref.\cite{Deng-Zhang-2018}.
Next, we have
\begin{equation}\label{vn+1C}
v_{n+1}=|A|\left(|\widehat{N};\widetilde{M},M+2|+|\widehat{N-1},N+1;\widetilde{M+1}|-|\widehat{N};\widetilde{M+1}|\right).
\end{equation}
Substituting them into Eq.\eqref{blf-3}, we get
\begin{equation}
\begin{array}{rl}
 & v_{n+1}f_n-v_nf_{n+1}+g_{n+1}h_n+g_nh_{n+1}\\
 =& |A|^{-1}\Bigl[|\widetilde{N+1};\widetilde{M-1},M+1| |\widehat{N};\widehat{M}|-|\widehat{N};\widehat{M-1},M+1|
|\widetilde{N+1};\widehat{M}|\\
& ~~~~~~~ -|\widehat{N+1};\widehat{M-1}||\widetilde{N};\widehat{M+1}|\Bigr]\\
&+|A|\Bigl[|\widehat{N-1},N+1;\widetilde{M+1}||\widehat{N};\widehat{M}|
-|\widehat{N-1},N+1;\widehat{M}||\widehat{N};\widetilde{M+1}|\\
& ~~~~~~~ -|\widehat{N+1};\widetilde{M}||\widehat{N-1};\widehat{M+1}|\Bigr],
\end{array}
\end{equation}
which is null by employing Lemma \ref{lem1}.

Thus we complete the proof.
\end{proof}
\end{appendix}


\end{document}